\documentclass[11pt,a4paper]{article}
\usepackage{jheppub}
\usepackage{graphicx}
\usepackage{dcolumn}
\usepackage{arydshln}
\usepackage{amsmath}
\usepackage{bm}
\usepackage{yhmath}
\usepackage{textcomp}
\usepackage{epstopdf}
\usepackage{multirow}

\usepackage{mathrsfs}

\usepackage{txfonts}
\usepackage{subfig}

\usepackage{color}%please remove it after revising
\usepackage{ulem}%please remove it after revising

\allowdisplaybreaks

\title{A Missing Partner Model With $24$-plet Breaking SU(5)}
\author[]{Da-Xin Zhang}
\author[]{and Jun-hui Zheng}
\affiliation[]{School of Physics
and State Key Laboratory of Nuclear Physics and Technology,\\
Peking University, Beijing 100871, China}
\emailAdd{dxzhang@pku.edu.cn}
\emailAdd{jhzheng@pku.edu.cn}
\abstract{We give a  missing partner model  using $24$-plet instead of $75$-plet to break the SU(5) symmetry.
Fermion masses and mixing are generated through the Georgi-Jarlskog mechanism. The model is constructed at
 renormalizable level at very high energy. The perturbative region is extended for the unification gauge coupling.
Constrains by proton decay is also satisfied.}

\begin{document}

\maketitle

\pagenumbering{arabic}

\section{Introduction}\label{a}

Supersymmetry (SUSY) is used to solve the naturalness problem in Grand Unified Theory (GUT)\cite{georgi1974}.
In the SUSY GUT (SGUT) model of SU(5)\cite{georgi1981}, in addition to a $24$-plet  Higgs used to break the GUT symmetry, there are
a pair of Higgs superfields in $5+\overline 5$ which contain a pair of weak doublets   to break SU(2)$_L \times$ U(1)$_Y$ at the weak scale and to give masses to the Standard Model (SM) particles.
There exists the so-called  doublet-triplet splitting (DTS) problem. The weak Higgs doublets in the $5+\overline 5$
are required to be light ($\sim10^2$ GeV), while the color-triplets in the same representations must be sufficiently heavy ($> 10^{17}$ GeV) to suppress proton decay. The DTS problem can be solved in the missing partner model (MPM)\cite{Yanagida1982,Grinstein1982} by introducing  extra Higgs in  $50+\overline{50}$ and using a $75$-plet, instead of the $24$-plet, to break the GUT symmetry.

Another problem in the SGUT is that it predicts $m_{d}=m_{l}$ at the GUT scale, which is inconsistent with the data. The Frogatt-Nielsen mechanism (FNM)\cite{FNM} and Georgi-Jarlskog mechanism (GJM)\cite{georgi1979} are the commonly used approaches to generate correct fermion masses. Unlike the FNM using higher-dimensional operators, in the SUSY version of  GJM\cite{wuzhang}, an extra pair of Higgs in $45+\overline{45}$ are introduced to couple with the matter fields to generate correct masses and mixing renormalizably.

We have tried in \cite{zheng2012}  combining the MPM and GJM to construct a renormalizable model of SUSY SU(5).
As in the original MPM, the $75$-plet Higgs is used to break the GUT symmetry. The mass relations between the down-quarks and the charged-leptons are corrected by coupling these matter fields with both $\overline 5$ and  $\overline{45}$ through GJM. Two U(1) symmetries are introduced to separate the Higgs spectrum to realize gauge coupling unification through threshold effects. These U(1) symmetries are also used in forbidding the unwanted couplings of $10_F10_F45$ which make the prediction
 on proton decay uncontrollable. However, as large representations $75$ and $45+\overline{45}$ are all introduced at around the GUT scale,  the GUT gauge coupling has huge change in its $\beta$-function. Consequently, this coupling runs into the non-perturbative region far below the (reduced) Planck scale of $M_{Pl}\sim 2.4\times 10^{18}$GeV.

In this work, we will further improve the MPM by using $24$ instead of $75$ to break the SU(5).
It was noticed in  \cite{Berezhiani} that the product of two $24$s can act as an effective $75$.
We note that  this effective $75$  can be constructed   by the mediation of  a $45+\overline{45}$ pair at the renormalizable level.
%However, without using the GJM, the fermion sector  is too complicated.
We will also carry out
a full analysis of the model which was not done in the literature\cite{Berezhiani}.

The paper is organized as follows. A brief review on the previous MPM studies  is presented in Section \ref{aa}. We will  present the requirements of a realistic model in Section \ref{b0}. In this Section we will construct the MPM with $24$ first in \ref{b}, give a comprehensive analysis on the necessity of the double-MPM in \ref{c}, then illustrate how to realize the Double MPM using  U(1) symmetries in \ref{cc}, and realize the MPM in the presence of  $45+\overline{45}$ at the GUT scale in \ref{dd}. In Section \ref{d}, we will construct a  realistic model explicitly. Higgs spectrum will also be given. We will study the constraints on the parameters imposed by gauge coupling unification in Section \ref{e},  and carry out the proton decay study in Section \ref{f}. Finally in Section \ref{g} we will summarize.

\section{Review of the Previous MPMs}\label{aa}

In this section, we give a short review on the main results on the various MPMs . We will describe the content of these models, point out their successes and shortcomings which are to guide us building a realistic model in the present work.

\textit{\bf The Minimal  MPM} \quad

\begin{figure}
       \centering
       \includegraphics[scale=0.8]{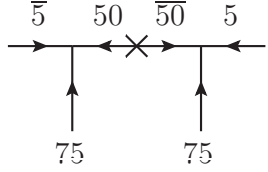}
       \caption{The generation of color-triplet masses in $5+\overline 5$ in the minimal MPM. A cross represents a VEV  of  the SU(5) singlet $1$. }\label{1.1}
\end{figure}

In the original version of the MPM, the minimal MPM\cite{Yanagida1982,Grinstein1982}, the Higgs superfields $5+\overline 5$ give the fermion  masses through
\begin{equation}\label{t2}
    W_F=\sqrt{2} f_{ij}~ \phi_i \cdot \psi_j \cdot \overline{5}+ \frac{1}{4} h_{ij}~ \psi_i \cdot \psi_j \cdot 5,
\end{equation}
where $\phi$'s and $\psi$'s are the $\overline{5}$- and $10$-matter superfields, respectively, and $i,j$ are the generation indices. The DTS problem is solved through the superpotential
\begin{equation}\label{t1}
    W= a~ \overline5 \cdot 75 \cdot 50 + b~ \overline{50} \cdot 75 \cdot 5 + c~ 1 \cdot \overline{50} \cdot 50,
\end{equation}
where a U(1) symmetry is introduced and the U(1) charges of the superfields are arranged to guarantee the absence of the term $(1\cdot )5\cdot \overline 5$. The singlet $1$ is used to break the U(1) symmetry and to give masses to all the components of the $50$ and $\overline{50}$. The $50+\overline{50}$ contain no weak doublet so that the Higgs doublet superfields of the MSSM are massless at the GUT scale $\Lambda$.
The mechanism of generating the color-triplet masses in $5+\overline 5$ can be seen in  Fig.\ref{1.1}.
Note that without
this U(1) symmetry additional baryon and lepton  number violation will occur through  $\psi_i\cdot\psi_j\cdot{50}$ whose coefficients are undetermined by the fermion masses.

It is the  effective triplet mass (ETM)
\begin{equation}\label{etm1}
    M_{ETM}\sim \frac{A^2}{N}
\end{equation}
which determines the
proton decay rates through the couplings in (\ref{t2}), where $A$ is the vacuum expecting value (VEV) of the $75$ to break the SU(5), and $N$ is the the VEV of $1$ to break the  U(1). There is a conflict in (\ref{etm1}): $N< <A\sim \Lambda$ is required to generate a rational ETM to suppress proton decay; however, the masses of $50+\overline{50}$, which are proportional to $N$, must be sufficient large ($> >\Lambda$) to keep the unified theory remain perturbative well above  the GUT scale.

 In \cite{alteralli2000}, the U(1) symmetry is chosen to break at the GUT scale.
% To enhance the  ETM to  around $6\times 10^{17}$ GeV,.
% the  SUSY breaking scale is tuned to be around $250$ GeV in accord with the requirement of gauge coupling
% unification.
The presence of $50+\overline{50}$ and $75$ at the  GUT scale leads the SU(5) gauge coupling to be
non-perturbative at around $10^{17}$ GeV.

\textit{\bf The Double MPM} \quad

To solve the conflict of the minimal MPM in (\ref{etm1}), the Double MPM  \cite{Hisano1995} was introduced where the MPM is used twice by the following settings. First, an extra pair of $5^\prime+\overline 5^\prime$ and an extra pair of $50^\prime+\overline{50}^\prime$ are introduced without couplings to the matter fields. Second, the MPM is applied separately in the two sectors with
$5 + 50 +\overline{50}+\overline 5^\prime$ and with $5^\prime +50^\prime+\overline{50}^\prime +\overline 5$. All the large
representations other than $75$ are put to be as heavy as $M_{Pl}$, resulting the triplets in $5+\overline 5^\prime$ and $5^\prime+\overline 5$ with masses $m_T$ and $m_T^\prime$ of the order $A^2/M_{Pl}$, while all the doublets are massless at this stage. Third, a perturbation of
$1\cdot5^\prime\cdot\overline 5^\prime$ is added, where the singlet $1$ gets a VEV$\sim 10^{11}$GeV to break the U(1) symmetry
which is used in realizing the two MPMs. Then below the GUT scale the superpotential for the Higgs sector is
\begin{equation}\label{t3}
    W = m_T~\overline{T}_{\overline{5}^\prime}\cdot T_5 +  m'_T~\overline{T}_{\overline{5}}\cdot T_{5^\prime}
    + c ~1 \cdot (\overline{T}_{\overline{5}^\prime}\cdot T_{5^\prime}+\overline{H}_{\overline{5}^\prime}\cdot H_{5^\prime}),
\end{equation}
where the doublets $\overline{H}_{\overline{5}^\prime}\cdot H_{5^\prime}$ in $5^\prime+\overline 5^\prime$  get masses  $c \langle 1 \rangle$ while those in $5+\overline 5$  are still massless. Taking the couplings with the matter fields the same as those in (\ref{t2}), the resultant ETM is now $\frac{m_T m_T'}{c N}>>\Lambda$ (N=$\langle 1 \rangle$), large enough to suppress proton decay. Furthermore, this large ETM is consistent with the threshold effects in unifying the gauge couplings\cite{Hisano1995}.
However, as was pointed out in \cite{Berezhiani}, there are still some allowed terms ($5\cdot\overline 5^\prime$ and $5^\prime\cdot\overline 5$) omitted `by hand' in the Double MPM Model.

In \cite{Berezhiani} the U(1) symmetry is chosen to break at a high scale $\sim 10^{17}$GeV instead of at an intermediate scale $\sim 10^{11}$GeV used in \cite{Hisano1995}. The omitted terms are now forbidden at the price that the last term in (\ref{t3}) is absent. Consequently, there will be two pairs of massless Higgs doublets at low energy, inducing large flavor changing neutral interactions at tree-level. To generate the masses for the extra doublets, non-renormalizale operators  were also used which, however, implies that the model is incomplete in the Higgs sector.

An important point observed in \cite{Berezhiani} is that an effective operator $\frac{(24\cdot 24)_{75}}{M_{Pl}}$
can act as an effective $75$ to realize the MPM.
Then the model remains to be perturbative far above the GUT scale.
Again, the  use of  non-renormalizale operator $\frac{(24\cdot 24)_{75}}{M_{Pl}}$  means that the Higgs sector is incomplete, which needs an explicit construction.

\textit{\bf The MPM with GJM} \quad

In the MPM it is a problem on how to describe the fermion masses and mixing. Usually the FNM was used when a high scale breaking U(1) is used. The fermion mass hierarchies and the mixing angles are attributed to powers of  $\frac{\Lambda}{M_{Pl}}$.
However, it is still difficult to avoid the unrealistic  relations $m_{d_i}=m_{e_i}$ \cite{Berezhiani,alteralli2000}.

In \cite{zheng2012}, we have applied the GJM together with the MPM. The Higgs superfields in ${45}+\overline{45}$ are introduced at the GUT scale to describe the fermion masses and mixing through
  \begin{equation}\label{t2t}
    W_{GJM}=\sqrt{2} f_{1ij}~ \phi_i \cdot \psi_j \cdot \overline{5}+ \sqrt{2} f_{2ij}~ \phi_i \cdot \psi_j \cdot \overline{45}+\frac{1}{4} h_{ij}~ \psi_i \cdot \psi_j \cdot 5.
\end{equation}
We have introduced two U(1) symmetries to forbid the unwanted terms in  $W_{GJM}$ and in the Higgs superpotential. One of
the U(1)s is broken at a high scale as in \cite{Berezhiani},  the other  is broken at an intermediate scale to give masses to the extra doublets. The main drawback of \cite{zheng2012} is that
 as the ${45}+\overline{45}$ are introduced, the SU(5) gauge coupling runs into the non-perturbative region just above the GUT scale. By using $24$ instead of $75$ in this work, one can hope to slow down the running  of the SU(5) gauge coupling.

\section{The Building Blocks of the Realistic Model}\label{b0}

In the MPM of SU(5) SGUT broken by $24$  instead of $75$, we need to construct the renormalizable model explicitly.
The problems in the minimal MPM need to be improved using the double MPM.
The double MPM will be realized by introducing two U(1) symmetries.
Using GJM, $45+\overline{45}$ at GUT scale need to be added to account for the fermion masses and mixing.

\subsection{The MPM with 24-plet}\label{b}

\begin{figure}
       \centering
       \includegraphics[scale=0.8]{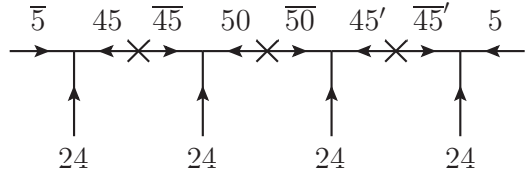}
       \caption{The generation of color-triplet masses in MPM with 24. A cross represents a O$(M_{Pl})$ mass.}\label{1.1b}
\end{figure}

 In the MPM, the $50+\overline{50}$ contain no doublet and act as filters preventing the doublets of $5+\overline{5}$
 from generating masses through the coupling $5 \cdot \overline{50} \cdot 75$ or $50 \cdot \overline{5} \cdot 75$. As was noticed in \cite{Berezhiani}, there can be an effective non-renormalizable operator $\frac{(24 \cdot 24)_{75}}{M_{Pl}}$ in the absence of the $75$. We note that this can be realized explicitly by introducing two pairs of $45+\overline{45}$  to generate the effective couplings with $\overline{5}$-$50$ and $\overline{50}$-$5$, respectively, which is illustrated in Fig.\ref{1.1b}.

The corresponding superpotential is
\begin{equation}\label{2.1}
\begin{split}
    W=&\frac{\sqrt{30}}{5}~a~\overline{5}_i 24_j^k 45^{ij}_{k} +
    \frac{M}{2}\overline{45}_{ij}^k 45^{ij}_{k}+
    \frac{\sqrt{30}}{5}~b~\overline{45}_{ij}^l 24_k^m 50^{ijk}_{lm}+
    \frac{N}{12}\overline{50}_{ijk}^{lm} 50^{ijk}_{lm}\\&\!\!
    +\frac{\sqrt{30}}{5}~c~\overline{50}_{ijk}^{lm} 24^i_l {45'}^{jk}_{m}+
    \frac{M'}{2}\overline{45}_{ij}^{\prime ~k} {45'}^{ij}_{k}+
    \frac{\sqrt{30}}{5}~d~\overline{45}_{ij}^{\prime ~k} 24^i_k 5^{j},
\end{split}
\end{equation}
where the coefficients are chosen for later convenience and $a,~b,~c,~d$ are all of order $1$. Some U(1) symmetries should be introduced to assure that the superpotential in (\ref{2.1}) is the most general form.
Differing in situations, the masses of $45+\overline{45}$, $45'+\overline{45}^\prime$ and $50+\overline{50}$ are generated from bilinear terms, or from triple couplings with the  $24$-plet or singlet which has a VEV.

It is necessary to emphasize that these two pairs of $45+\overline{45}$ must be different. The $45$ and $\overline{45}$ must have different U(1) charges from those of $45'$ and $\overline{45}'$, respectively; otherwise, the two couplings  with the filters $50+\overline{50}$ are not necessary, leading the doublets to be massive.  It is also necessary to emphasize that the superpotential given in (\ref{2.1}) is the only way to realize the  MPM,
any new bilinear term added additionally is forbidden. This can be seen from Fig. \ref{1.1b}, the presence of any  term of $\overline{5}$-$5$, $\overline{45}$-$45'$ and $45$-$\overline{45}'$ will introduce a  graph connecting $5$ and $\overline{5}$ without the filters $50$-$\overline{50}$, which generates doublet masses and thus destroys MPM. Trilinear terms $\overline{5}$-24-$45'$, $\overline{45}$-24-$5$, $50$-24-$\overline{45}'$ and $45$-24-$\overline{50}$ can not emerge because $45$ and $\overline{45}$  have different quantum numbers with those of $45'$ and $\overline{45}'$.

When the SM singlet of $24$ obtains a VEV
\begin{equation}\label{2.2a}
    \langle24\rangle=\frac{A}{\sqrt{30}}diag(2,2,2,-3,-3),
\end{equation}
it gives the  mass matrix for the color triplets
\begin{equation}\label{2.2}
  M_T= \begin{array}{c|ccccc}
             & T_5 & T_{45} & T_{45'} & T_{50}\\
           \hline
           \overline{T}_{\overline{5}} & 0 &- a A & 0 & 0 \\
           \overline{T}_{\overline{45}} & 0 & M &  0  &-\frac{4}{\sqrt{3}} b A  \\
           \overline{T}_{\overline{45}'} &- d A & 0  & M' & 0 \\
           \overline{T}_{\overline{50}} & 0 &  0 & -\frac{4}{\sqrt{3}} c A & N,
         \end{array}
\end{equation}
and the  mass matrix for the weak doublets
\begin{equation}\label{2.3}
    M_D=\begin{array}{c|ccc}
             & H_5 & H_{45} & H_{45'} \\
           \hline
           \overline{H}_{\overline{5}} & 0 &-\frac{\sqrt{3}}{2} a A & 0\\
           \overline{H}_{\overline{45}} & 0 & M & 0 \\
           \overline{H}_{\overline{45}'} & -\frac{\sqrt{3}}{2} d A & 0 & M'.
         \end{array}
\end{equation}
They suggest that a pair of massless doublets exist while all the triplets are massive, so that the DTS
is fulfilled. If only the $5+\overline{5}$ couple to matter fields as in (\ref{t2}), the ETM is
\begin{equation}\label{2.4}
 M_{ETM}= \big[({M_T}^{-1})_{11}\big]^{-1}=-\frac{16 a b c d A^4}{3 MM'N},
\end{equation}
which is consistent with the result given in the non-renormalizable model\cite{Berezhiani}.

The effective MPM with $24$ replacing $75$ can be seen if we put $M$ and $M'$ at the highest scale of the model. The $45+\overline{45}$s are heavy so that they are integrated out above the GUT scale, resulting the effective triplet mass matrix as
\begin{equation}\label{2.5}
 \begin{array}{c|cc}
             & T_5 & T_{50}\\
           \hline
           \overline{T}_{\overline{5}} & 0 & -\frac{4}{\sqrt{3}} \frac{ab A^2}{M}  \\
           \overline{T}_{\overline{50}} & -\frac{4}{\sqrt{3}} \frac{cd A^2}{M'} &  N,
         \end{array}
\end{equation}
which suggests a combination of two $24$, i.e., $\frac{1}{M}~24\cdot24$ acts an effective $75$-plet to realize the couplings with $5$-$\overline{50}$ and $\overline{5}$-$50$. The effective superpotential becomes
\begin{equation}\label{2.6}
    W^e=-\frac{12a b}{5M}~\overline{5}_i (24_j^l 24_k^m)_{75} 50^{ijk}_{lm}+
    \frac{N}{12}\overline{50}_{ijk}^{lm} 50^{ijk}_{lm}-
    \frac{12c d}{5M'}~\overline{50}_{ijk}^{lm} (24^k_m 24^j_l)_{75} 5^{i},
\end{equation}
just mimicking that of the original MPM (\ref{t1}). Note that
the ETM given in E.q. (\ref{2.4}), which can be also read off from (\ref{2.6}), is always too small to suppress proton decay  even the $50+\overline{50}$ have masses of the order of  the GUT scale.
%In practice they have masses of the Planck scale making the ETM even smaller,
%which worsens the situation on proton lifetime.

\subsection{Requirement of the Double MPM}\label{c}

As we have shown that, if we keep all large representations to be at Planck scale while leaving one pair of  $5+\overline{5}$ and the $24$-plet at the GUT scale, the ETM will be too small. In the Double MPM with $75$\cite{Hisano1995} presented in Section \ref{aa}, an extra pair of $5^\prime+\overline{5}^\prime$ at an intermediate scale can fix the problem.
Here we discuss in details the possible forms of the doublet and triplet mass matrices,
and then show how to use MPM in a realistic model.

First, we consider the doublet sector. There are one pair of massless doublets $H_u$ and $H_d$ in the MSSM which couple to the matter fields. Taking at the GUT scale the superpotential for the matter sector as (\ref{t2}),  $H_u$ and $H_d$ must come mainly from $5$ and $\overline 5$, respectively,
which requires the mass matrix for the doublets to be
\begin{equation}\label{3.1}
        \begin{array}{c|cc}
      0 & H_5 & H_{5'} \\
      \hline
      \overline{H}_{\overline{5}} & 0 & m_1 \\
      \overline{H}_{\overline{5}'} & 0 & m_2
    \end{array}~,
\end{equation}
or
 \begin{equation}\label{3.1b}
    \begin{array}{c|cc}
      0 & H_5 & H_{5'} \\
      \hline
      \overline{H}_{\overline{5}} & 0 & 0\\
      \overline{H}_{\overline{5}'} & m_1 & m_2
    \end{array}~,
\end{equation}
if no fine-tuning exists.
Here one of $m_1$ and $m_2$ can be 0. We have
\begin{equation}\label{m1m2}
   m_1 \lesssim m_2,
\end{equation}
otherwise some fermions cannot get masses through the couplings in (\ref{t2}).
We will take the form in (\ref{3.1}) as the example, while  the other form in (\ref{3.1b}) follows the same discussion.

Secondly, we consider the corresponding mass matrix for the triplets. Because all triplets are massive, the triplet mass matrix must be one of the following three forms
\begin{equation}\label{3.2}
    \begin{array}{c|cc}
        & T_5 & T_{5'} \\
      \hline
      \overline{T}_{\overline{5}} & 0 & m'_1 \\
      \overline{T}_{\overline{5}'} & \overline{m} & m'_2
    \end{array}~,~~~~~
    \begin{array}{c|cc}
        & T_5 & T_{5'} \\
      \hline
      \overline{T}_{\overline{5}} & \overline{m}' & m'_1 \\
      \overline{T}_{\overline{5}'} & \overline{m} & m'_2
    \end{array}~,~~~~~
    \begin{array}{c|cc}
        & T_5 & T_{5'} \\
      \hline
      \overline{T}_{\overline{5}} & \overline{m}' & m'_1 \\
      \overline{T}_{\overline{5}'} & 0 & m'_2
    \end{array}~,
\end{equation}
where the $\overline{m}'$ and $\overline{m}$ must come from the mechanism of the MPM because there is no corresponding mass term  $\overline{H}_{\overline{5}}\cdot H_5$ or $\overline{H}_{\overline{5}'}\cdot H_{5}$, as can be seen from ({\ref{3.1}). The ETMs are
\begin{equation}\label{3.3}
    -\overline{m}\frac{m'_1}{m'_2}, ~~~~~\overline{m}'-\overline{m}\frac{m'_1}{m'_2},~~~~~\overline{m}'
\end{equation}
for these three cases, respectively. Being ETMs themselves which have been shown in (\ref{2.4}), $\overline{m}$ and $\overline{m}'$ are of the order $\frac{\Lambda^4}{M_{Pl}^3} \ll \Lambda$.
The third value ($\overline{m}'$) in (\ref{3.3}) is too small compared to the requirement of a large ETM, thus we exclude the last form in (\ref{3.2}). The other two ETMs are possibly large enough and are approximately equal, providing
\begin{equation}\label{m1pm2p}
    m'_2\ll m'_1.
\end{equation}
It suggests that we need not  differ  the first two forms in (\ref{3.2}).

A third relation among the mass parameters folows that without fine-tuning we have
\begin{equation}\label{m2m2p}
  m_2 \Bigg(\begin{array}{c}
    \sim
 \\ \ll \end{array}\Bigg) m_2'
  \end{equation}
and
 \begin{equation}\label{m1m1p}
  m_1 \Bigg(\begin{array}{c}
    \sim
 \\ \ll \end{array}\Bigg) m_1',
  \end{equation}
where the lower possibilities follows if $m_{1,2}'$ is generated through a MPM.

(\ref{m1m2}), (\ref{m1pm2p}) and (\ref{m2m2p}) together give
\begin{equation}\label{mall}
  m_1\lesssim m_2
  \Bigg(\begin{array}{c}
    \sim
 \\ \ll \end{array}\Bigg)
  m_2' \ll m_1'.
  \end{equation}
Comparing with (\ref{m1m1p}), this proves that $m_1'$ is generated though a MPM. The full relations (\ref{mall}) also prove that in the simplest case
of putting $m_1\sim 0$, there exists at least an intermediate scale for $m_2$ and $m_2'$.

To summarize, the mass matrices of doublets and triplets are
\begin{equation}\label{3.5}
     \begin{array}{c|cc}
        & H_5 & H_{5'} \\
      \hline
      \overline{H}_{\overline{5}} & 0 & ~0 \\
      \overline{H}_{\overline{5}'} & 0 & ~O(m_2)
    \end{array}~,~~~~~{\rm and}~~~
     \begin{array}{c|cc}
        & T_5 & T_{5'} \\
      \hline
      \overline{T}_{\overline{5}} & 0, ~O(\overline m) & ~O({\overline{m}}) \\
      \overline{T}_{\overline{5}'} & O(\overline{m}) & ~O(m_2)
    \end{array}~,
\end{equation}
respectively. The Double MPM is needed which is depicted in Fig. \ref{3.6}.
It can be noted that the discussions above  apply also in the model with 75-plet\cite{Hisano1995}.
\begin{figure}
       \centering
       \includegraphics[scale=0.85]{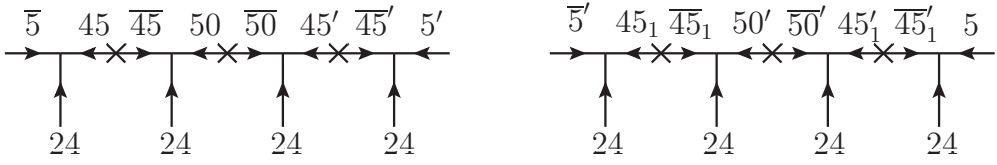}
       \caption{The Double MPM where the MPM is used in two sectors separately. A cross represents a O$(M_{P})$ mass.
       %These two graphs are connected through $1\cdot 5^\prime\cdot \overline 5^\prime$.
       }\label{3.6}
\end{figure}

\subsection{Two U(1)s}\label{cc}

Now  a realistic model has  several scales in the Higgs sector. The highest scale is the Planck scale $M_{Pl}$
of all the large representations; a GUT breaking scale $A=\langle 24 \rangle$; a MPM-generated scale $\overline{m}\sim\frac{A^4}{M_{Pl}^3}$; an intermediate
scale $m_2\ll \overline{m}$; and a SUSY or electro-weak scale which is set to zero for simplicity.

The simplest approach to introduce the Planck scale masses
in the superpotential is to introduce a U(1)$_P$ symmetry which is broken by a SU(5) singlet $P$ whose VEV is $ \langle P \rangle=O(M_{Pl})$.

To generate the intermediate scale, we can simply introduce another U(1)$_S$ symmetry which is broken by a SU(5) singlet $S$ whose VEV is $ \langle S \rangle\ll O(\frac{A^4}{M_{Pl}^3})$.
Coupling of $S$ with $5^\prime$-$\overline{5}^\prime$ generates the $O(m_2)$ entries in (\ref{3.5}).
These two U(1)s are also needed in avoiding unwanted large masses in the general superpotential of a realistic model.

\subsection{The MPM and the GJM}\label{dd}

In the renormalizable SU(5) models, the GJM is used  to give the down-quark and charged-lepton masses
by introducing $\overline{45}$ which also couples to $\overline 5_F$-$10_F$. In the SUSY version of GJM, this can also be realized by using the MPM in the presence of $75$\cite{zheng2012}.

In the present case of using $24$ instead of $75$ to break GUT symmetry, we note that the presence of the coupling $\overline{45}$-50-24 makes this realization easier.
% To realize MPM, the  superpotential is
% \begin{equation}\label{45a}
%     W_{45}=\overline{45}\cdot 24\cdot 50+\overline{50} \cdot 24 \cdot 5 +  1 \cdot \overline{50} \cdot 50,
% \end{equation}
Coming back to the superpotential (\ref{2.1}), the parameter $M$ is now set to be at the scale $\Lambda$ instead of at  $M_{Pl}$.
Consequently, in realizing MPM in Fig. 2, the $45$-$\overline{45}$  are not integrated out above the GUT scale, at which the Higgs sector now contains $24$, $5+\overline 5$ and $45+\overline{45}$. The mass matrix for the triplets is
\begin{equation}\label{3.21}
  M_T= \begin{array}{c|ccc}
             & T_5 & T_{45} \\
           \hline
           \overline{T}_{\overline{5}} & ~0 & ~-a A  \\
           \overline{T}_{\overline{45}} & -\frac{16 b c d A^3}{3 M'N} & ~\epsilon A \\
         \end{array}
\end{equation}
and that for the doublets is
\begin{equation}\label{3.31}
    M_D=\begin{array}{c|cc}
             & H_5 & H_{45}  \\
           \hline
           \overline{H}_{\overline{5}} & 0 &-\frac{\sqrt{3}}{2} a A \\
           \overline{H}_{\overline{45}} & 0 & \epsilon A .\\
         \end{array}
\end{equation}
Here $\epsilon$ is of $O(1)$. There are still a pair of massless Higgs doublets.
Those dimension-5 operators of proton decay mediated by the $\overline{5}-5$ and $\overline{45}-5$ triplets are proportional to
\begin{equation}\label{3.32}
    \Big(M_T^{-1}\Big)_{11}= -\frac{3 \epsilon M'N}{16 a b c d A^3} , ~~~{\rm and} ~~~\Big(M_T^{-1}\Big)_{12}=-\frac{3 M'N}{16 b c d A^3},
\end{equation}
respectively, so the ETM is determined by these two quantities which is
further enhanced by $\frac{M_{Pl}}{A}$ compared to that in (\ref{2.4}).

%As the double-MPM is used as in \subsection{c}, we will not apply the GJM in the sector to couple with

\section{The Realistic Model and the Higgs Spectrum}\label{d}

 We are now ready to construct a realistic SUSY SU(5) model with $24$-plet to breaking GUT symmetry. The
requirements are the following.
\begin{itemize}
  \item MPM for the $\overline 5+ 45+ \overline{45}+5^\prime$ sector, which requires the heavy   fields  $50+\overline{50}$ and $45^\prime+\overline{45}^\prime$ at $M_{Pl}$;
  \item MPM for the $\overline 5^\prime+5$ sector, which requires the heavy   fields  $50^\prime+\overline{50}^\prime$, $45_1+\overline{45}_1$  and $45_1^\prime+\overline{45}_1^\prime$ at $M_{Pl}$;
  \item A U(1)$_P$ symmetry with a singlet $P$ whose VEV is of the order  $M_{Pl}$;
  \item A U(1)$_S$ symmetry with a singlet $S$ whose VEV is at an intermediate scale. $S$ also gives masses to $5^\prime+\overline 5^\prime$;
  \item A direct coupling $\overline 5\cdot 24\cdot 45$  and a mass term $45~\overline{45}$ at $\Lambda$;
  \item Absence of all other couplings which spoil the MPMs. This is realized by arranging  appropriate U(1) charges.
\end{itemize}
These requirements are depicted in Fig. \ref{3}.
% Now we ow to use the MPM in a realistic model. As was shown in previous, both the masses of  $\overline{T}_{\overline{5}}$-$T_{5'}$ and $\overline{T}_{\overline{5}'}$-$ T_5$ come from MPM, so MPM must be used twice at least. Two $50+\overline{50}$s and four $45+\overline{45}$s are needed, and the corresponding diagram is shown in Fig.\ref{3.6}. A toy model contains just all couplings emerging in Fig.\ref{3.6} and a trilinear coupling $1_s \cdot \overline{5}' \cdot 5'$ which connects these two diagrams. All large representation Higgs are heavy as $M_p$ and only $5+\overline{5}$ couples to matter fields. By integrating out heavy particles $45+\overline{45}$s and $50+\overline{50}$s, the effective mass terms at $\Lambda$ is $W_m = O(\Delta)~\overline{H}_{\overline{5'}}\cdot H_{5'}+O(\overline{m})~\overline{T}_{\overline{5'}}\cdot T_5 + O(\overline{m})~\overline{T}_{\overline{5}}\cdot T_{5'}+ O(\Delta)~\overline{T}_{\overline{5'}}\cdot T_{5'}$.
\begin{figure}
       \centering
       \includegraphics[scale=0.85]{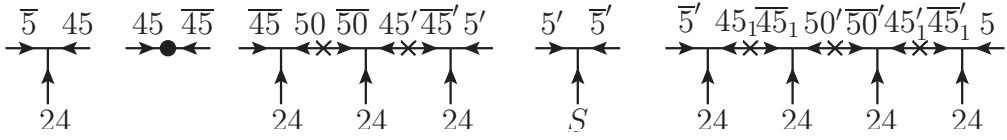}
       \caption{The realistic model. A dot stands for a mass of $O(\Lambda)$, while a cross represents a coupling with $P$ which has a VEV of $O(M_{Pl})$ .}\label{3}
\end{figure}

We assign the $U(1)_S$ and $U(1)_P$ charges for these Higgs multiplets as in Table \ref{4.1}.
\begin{table}
\begin{center}
  \begin{tabular}{|c|c|c|c|c|c|c|c|c|c|c|c|c|c|c|c|c|c|c|c|c|}
   \hline
    & $\overline{5}$ & $45$ & $\overline{45}$ & $50$ & $\overline{50}$ & $45'$ & $\overline{45}'$ & $5'$&
    $\overline{5}'$ & $ 45_1$ & $ \overline{45}_1$  & $ 50'$ & $ \overline{50}'$  & $ 45'_1$& $\overline{45}'_1$ & $5$ \\
   \hline
    $U(1)_S$ & $1$ & $-1$ & $1$ & $-1$ & $1$ & $-1$ & $1$ & $-1$ & $0$ & $ 0$ & $0$  & $ 0$ & $0$ & $0$ & $0$& $0$ \\
   \hline
     $U(1)_P$ & $5$ & $-5$ & $5$ & $-5$ & $4$ & $-4$ & $3$ & $-3$ & $3$ & $ -3$ & $ 2$  & $ -2$ & $1$ & $-1$ & $0$& $0$ \\
   \hline
   \end{tabular}
   \caption{$U(1)_S$ and $U(1)_P$ charges for the Higgs superfields.}\label{4.1}
\end{center}
\end{table}
%Two singlet $S$ and $P$ which are used to break $U_s(1)$ and $U_p(1)$ have quantum numbers $(1,0)$ and $(0,1)$ respectively, while the 24 which is used to break SU(5) to MSSM has $(0,0)$. $\overline{45}$ and $\overline{5}$ have same U(1) quantum numbers because they both couple to matter fields.
The most general renormalizabe  superpotential for the Higgs sector is
\begin{eqnarray}
      W&= &\lambda A 24^i_j 24^j_i + \frac{2\sqrt{30}}{3}\lambda24^i_j 24^j_k 24^k_i+\frac{\Delta}{\langle S \rangle}~S~\overline{5}^\prime_i 5^{\prime i}+\frac{\sqrt{30}}{5}~a~\overline{5}_i 24_j^k 45^{ij}_{k}\notag \\&&
        +\Big(\frac{M}{2}\overline{45}_{ij}^k 45^{ij}_{k}
        + \frac{z_1\sqrt{15}}{\sqrt{2}}\overline{45}_{ij}^k 24_k^{k'} 45^{ij}_{k'}+\frac{z_2 \sqrt{15}}{\sqrt{2}}\overline{45}_{ij}^k 24_{i'}^{i} 45^{i'j}_{k}\Big)
        +\frac{\sqrt{30}}{5}~b~\overline{45}_{ij}^l 24_k^m 50^{ijk}_{lm}\notag\\&&
        +\frac{N}{12\langle P \rangle}~P~\overline{50}_{ijk}^{lm} 50^{ijk}_{lm}
        +\frac{\sqrt{30}}{5}~c~\overline{50}_{ijk}^{lm} 24^i_l {45'}^{jk}_{m}
        +\frac{M'}{2\langle P \rangle}~P~\overline{45}_{ij}^{\prime k} {45'}^{ij}_{k}
        +\frac{\sqrt{30}}{5}~d~\overline{45}_{ij}^{\prime k} 24^i_k 5'^{j}\notag\\&&
        +\frac{\sqrt{30}}{5}~a'~ \overline{5}^\prime_i 24_j^k {45_1}^{ij}_{k}
        +\frac{M_1}{2 \langle P \rangle}~P~{\overline{45_1}}_{ij}^k {45_1}^{ij}_{k}
        +\frac{\sqrt{30}}{5}~b'~{\overline{45_1}}_{ij}^l 24_k^m {50'}^{ijk}_{lm}
        +\frac{N'}{12 \langle P \rangle}~P~\overline{50}_{ijk}^{\prime lm} {50'}^{ijk}_{lm}\notag\\&&
        +\frac{\sqrt{30}}{5}~c'~\overline{50}_{ijk}^{\prime lm} 24^i_l {45'_1}^{jk}_{m}
        +\frac{M'_1}{2 \langle P \rangle}~P~{\overline{45_1}}_{ij}^{\prime k} {45'_1}^{ij}_{k}
        +\frac{\sqrt{30}}{5}~d'~{\overline{45_1}}_{ij}^{\prime k} 24_k^i 5^{j},\label{4.2}
\end{eqnarray}
where the coefficients are chosen for later convenience and all the trilinear couplings are $O(1)$. The mechanism of breaking the U(1)s can be found in  e.g. \cite{Hisano1995}. We will not discuss the properties of these U(1)s which are irrelevant in main features of the present study.

Below the Planck scale, the $U(1)_P$ symmetry breaks when the SU(5) singlet $P(0,1)$ obtain a VEV $\langle P \rangle$.  This leads the $50, \overline{50}, 50',  \overline{50'},  45_1,  \overline{45_1},  45'_1, \overline{45'_1}$ and $45', \overline{45'}$ to be heavy. The SU(5) symmetry breaks when the $24$(0,0)obtains a VEV $A$ (see (\ref{2.2a})), while the $U(1)_S$ symmetry breaks at a lower scale when the SU(5) singlet $S(1,0)$ obtains a VEV $\langle S \rangle$.
The weak doublet mass matrix is
\begin{equation}\label{4.3}
    M_D= \begin{array}{c|ccc:ccc}
             & H_{5}& H_{45_1} & H_{45'_1} & H_{5'} & H_{45} & H_{45'}\\
           \hline
           \overline{H}_{\overline{5}} & 0 & 0 & 0 & 0 &-\frac{\sqrt{3}}{2} a A & 0 \\
           \overline{H}_{\overline{45}} & 0 & 0 & 0 & 0 & m & 0\\
           \overline{H}_{\overline{45}^\prime } & 0 & 0 & 0 & -\frac{\sqrt{3}}{2} d A & 0 & M'\\
           \hdashline
           \overline{H}_{\overline{5}^\prime} & 0 &-\frac{\sqrt{3}}{2}  a' A & 0  & \Delta & 0 & 0 \\
           \overline{H}_{\overline{45}_1} & 0 & M_1 & 0 & 0 & 0 & 0\\
           \overline{H}_{\overline{45}^\prime_1} & -\frac{\sqrt{3}}{2} d' A & 0 & M'_1 & 0 & 0 & 0,\\
         \end{array}
\end{equation}
and the color triplet mass matrix is
\begin{equation}\label{4.4}
   M_T=  \begin{array}{c|cccc:cccc}
            & T_{5}& T_{45_1} & T_{45'_1}& T_{50'} & T_{5'} & T_{45} & T_{45'}& T_{50} \\
           \hline
           \overline{T}_{\overline{5}} & 0 & 0 & 0 & 0& 0 & -a A & 0 &0 \\
           \overline{T}_{\overline{45}} & 0 & 0 & 0 & 0& 0 & \tilde{m} &0& -\frac{4}{\sqrt{3}}bA \\
           \overline{T}_{\overline{45}^\prime}& 0 & 0 & 0 & 0 & -dA & 0 & M'& 0\\
           \overline{T}_{\overline{50}} & 0 & 0 & 0 & 0& 0 & 0 & -\frac{4}{\sqrt{3}}cA & N\\
           \hdashline
           \overline{T}_{\overline{5}^\prime}  & 0 & -a'A& 0 & 0 & \Delta& 0 & 0 & 0\\
           \overline{T}_{\overline{45}_1} & 0 &M_1&0&-\frac{4}{\sqrt{3}}b'A& 0 & 0 & 0 & 0\\
           \overline{T}_{\overline{45}^\prime_1} & -d' A & 0 & M'_1 & 0& 0 & 0 & 0 & 0\\
           \overline{T}_{\overline{50}^\prime} & 0& 0 & -\frac{4}{\sqrt{3}}c'A & N'& 0 & 0 & 0 & ~0,\\
         \end{array}\\
\end{equation}
where
\begin{eqnarray}
m &=& M-\frac{7}{4}z_1A-\frac{19}{8}z_2A,\\
\tilde{m} &=& M-\frac{1}{2}z_1A+\frac{3}{4}z_2A .
\end{eqnarray}
The massless doublets corresponding to those in the MSSM are
\begin{eqnarray}
H_u&=& x_1 H_{5}+ x_2 H_{45_1} + x_3 H_{45'_1} + x_4 H_{5'} + x_5 H_{45} + x_6 H_{45'}, \\
H_d&=& y_1 \overline{H}_{\overline{5}} + y_2 \overline{H}_{\overline{45}}  + y_3 \overline{H}_{\overline{45}^\prime}
+ y_4 \overline{H}_{\overline{5}^\prime}+ y_5 \overline{H}_{\overline{45}_1} + y_6  \overline{H}_{\overline{45}^\prime_1},
\end{eqnarray}
satisfying $(M_D)_{ij} x_j =0$ and $y_i (M_D)_{ij}=0$. The corresponding solutions are
\begin{eqnarray}
      x&=&\frac{1}{\sqrt{\frac{3}{4}{d'}^2 A^2+ {M'_1}^2}}(M'_1,0 ,\frac{\sqrt{3}}{2} d' A,0,0,0),\label{4.5a}\\
      y&=&\frac{1}{\sqrt{\frac{3}{4}a^2 A^2+ m^2}}(m,\frac{\sqrt{3}}{2} a A,0,0,0,0).\label{4.5b}
\end{eqnarray}
In the present case $\langle P \rangle \gg A$, we can approximate $x$ as $x=(1,0,0,0,0,0)$.

At the GUT scale, by integrating out the all heavy fields,
the effective non-renormalizable superpotential is
\begin{eqnarray}
      W^{eff}= &&\lambda A 24^i_j 24^j_i + \frac{2\sqrt{30}}{3}\lambda24^i_j 24^j_k 24^k_i+\frac{\Delta}{\langle S \rangle}~S~\overline{5'}_i {5'}^{i}\notag\\&&
        -\frac{48 a'b'c'd'}{25M^{}_1M'_1N'}\Big({\overline{5'}}^{}_{[i}24^l_j 24^m_{k]}+\frac{1}{4}\delta^{[l}_{[i}{\overline{5'}}^{}_{[j}24^{m]}_k24^{s}_{(s)]]}
        +\frac{1}{6}\delta^l_{[i}\delta^m_j{\overline{5'}}^{}_{[k}24^{s}_{(s)} 24^{t}_{(t)]]}\Big)\cdot\notag\\&&
        \Big({5}^{[i}24_l^j 24_m^{k]}+\frac{1}{4}\delta_{[l}^{[i}{5}^{[j}24_{m]}^k24_{h}^{(h)]]}+\frac{1}{6}\delta_l^{[i}\delta_m^j{5}^{[k}24_{h}^{(h)} 24_{g}^{(g)]]}\Big)\notag\\&&
        +\frac{\sqrt{30}}{5}~a~\overline{5}_i 24_j^k 45^{ij}_{k}
        +\frac{M}{2}{\overline{45}}_{ij}^k 45^{ij}_{k}+ \frac{z_1\sqrt{15}}{\sqrt{2}}\overline{45}_{ij}^k 24_k^{k'} 45^{ij}_{k'}+\frac{z_2 \sqrt{15}}{\sqrt{2}}\overline{45}_{ij}^k 24_{i'}^{i} 45^{i'j}_{k}\notag\\&&
        +\frac{\sqrt{30}bcd}{25M'N}\Big({\overline{45}}^{[l}_{[ij} 24^{m]}_{k]}+\frac{1}{4}\delta^{[l}_{[i}{\overline{45}}^{[m}_{[jk}24^{(s)]]}_{(s)]]}
        +\frac{1}{6}\delta^l_{[i}\delta^m_j{\overline{45}}^{[s}_{[k(s)} 24^{t]}_{(t)]]}\Big)\cdot\notag\\&&
        \Big({5'}^{[i}24_l^j 24_m^{k]}+\frac{1}{4}\delta_{[l}^{[i}{5'}^{[j}24_{m]}^k24_{h}^{(h)]]}+\frac{1}{6}\delta_l^{[i}\delta_m^j{5'}^{[k}24_{h}^{(h)} 24_{g}^{(g)]]}\Big),\label{4.11}
\end{eqnarray}
where {\it e.g.}, $\delta_l^{[i}\delta_m^j{5}^{[k}24_{h}^{(h)} 24_{g}^{(g)]]}$ means to anti-symmetrize the up-indices $k,h,g$ first, and then to anti-symmetrize the $i,j,k$ ({\it i.e.} in this second step,  $h,g$ are not involved in the operation), while ${\overline{5}}^{\prime}_{[i}24^l_j 24^m_{k]}$ means to anti-symmetrize the down-indices $i,j,k$.
% The expressions in the brackets of (\ref{4.11}) act as an effective $50$ or $\overline{50}$s,
% so that the double-MPM is fulfilled.
The effective mass matrix for the weak doublets is
\begin{equation}\label{4.6}
    M^{eff}_D \sim \begin{array}{c|c:cc}
             & H_{5} & ~H_{45} & ~H_{5'}\\
           \hline
           \overline{H}_{\overline{5}} & 0 & ~-\frac{\sqrt{3}}{2} a A & ~0 \\
           \overline{H}_{\overline{45}} & 0 & ~m & ~0\\
           \hdashline
           \overline{H}_{\overline{5}^\prime} & 0 & ~0 & ~~\Delta, \\
         \end{array}
\end{equation}
from which one find two pairs of doublets with non-zero masses
\begin{equation}\label{4.7}
     M_+=\sqrt{m^2+\frac{3}{4}(a A)^2}, ~~
  M_-=\Delta,
\end{equation}
and a pair of massless doublets
\begin{equation}\label{4.8}
    H_u=H_5, ~~H_d= \frac{m}{M_+} \overline{H}_{\overline{5}} +  \frac{\sqrt{3} a A}{2 M_+}\overline{H}_{\overline{45}}.
\end{equation}
This  is consistent with (\ref{4.5a}) and (\ref{4.5b}). The  color triplets have the effective mass matrix
\begin{equation}\label{4.9}
    M^{eff}_T \sim \begin{array}{c|c:cc}
             & T_{5} & ~T_{45} & ~T_{5'}  \\
           \hline
           \overline{T}_{\overline{5}} & ~0 & ~-a A & 0 \\
           \overline{T}_{\overline{45}} & ~0 & \tilde{m} & ~ -\frac{16 b c d A^3}{3 M'N} \\
           \hdashline
           \overline{T}_{\overline{5}^\prime} & -\frac{16 a' b' c' d' A^4}{3 M^{}_1M'_1N'} &~ 0  &
           ~\Delta.
         \end{array}
\end{equation}
The effective mass matrices  (\ref{4.6}) and (\ref{4.9}) are extensions of  (\ref{3.5}), which can be seen from their block matrix forms.
The masses of other particles at $\Lambda$ are listed in Table \ref{3.11}. The spectrum is to be constrained by the requirement of gauge coupling unification through the threshold effects.
\begin{table}
  \centering
  \begin{tabular}{|c|c|c|c|}
    \hline
    Higgs superfield & Reps under SM groups & Mass  & Reps in $SU(5)$ \\
    \hline
    $ H^{a \alpha}_b , H_{a\alpha}^b $ & $ (8,2,\frac{1}{2}),(8,2,-\frac{1}{2}) $ & $ M+2z_1A-\frac{1}{2}z_2 A $
    & $45$,$\overline{45}$ \\
    \hline
    $ H^{a \alpha}, H_{a \alpha} $ & $ (3,2, \frac{7}{6}),(\overline{3},2, -\frac{7}{6}) $ & $ M-3z_1A+2z_2 A $
    & $45$,$\overline{45}$ \\
    \hline
    $ H_{ab(s)}, H^{ab}_{(s)} $ & $(\overline{6} ,1 ,-\frac{1}{3}),(6 ,1 , \frac{1}{3}) $ & $ M+2z_1A+2z_2 A  $ & $45$,$\overline{45} $ \\
    \hline
    $H^{a \alpha}_{\beta},H_{a \alpha}^{\beta}$ & $(3,3,-\frac{1}{3}),(\overline{3},3,\frac{1}{3})$ & $M-3z_1A-\frac{1}{2}z_2 A $ & $45$,$\overline{45}$ \\
    \hline
    $H^a,\overline{H}_a $ & $(3,1,-\frac{4}{3}), (\overline{3},1,\frac{4}{3})$ & $M+2z_1A-3z_2 A $ & $45$,$\overline{45}$ \\
    \hline
    $ \Sigma^{a}_b $ & $ (8,1,0)$ & $ 5 \lambda A $ & $ 24 $ \\
    \hline
    $\Sigma^a,\overline{\Sigma}_a $ & $(1,3,0)$ & $5\lambda A $ & $24$ \\
    \hline
    $ \Sigma_0 $ & $(1,1,0) $ & $\lambda A $ & $ 24 $ \\
    \hline
  \end{tabular}
  \caption{The spectrum of other Higgs superfields at the GUT scale. Here
  $a,b$ are color indexes, $\alpha,\beta$ are flavor indexes.}\label{3.11}
\end{table}
To illustrate numerically, we also give the spectrum in Table \ref{5.9} using a set of   representative parameters.
A pair of doublets exist at $1.0\times 10^{6}$ GeV, which is the $U(1)_S$ symmetry breaking scale. There exist also
two pairs of triplets with masses around $\sim 10^{11-13}$GeV below the GUT scale.
\begin{table}
  \centering
  \begin{tabular}{|c|c|}\hline
    Higgs multiplets & Masses (GeV)\\
    \hline
    $T^a ,\overline{T}_a$ &  \big($~1.3\times 10^{11},~1.3\times 10^{13},~2.5\times 10^{16}$\big)\\
    \hline
    $ H^{\alpha},\overline{H}_{\alpha} $ & \big($~0,~1 \times 10^6,~2.4\times 10^{16}$\big)\\
    \hline
    $ H^{a \alpha}_b , H_{a\alpha}^b $ & $ 1 \times 10^{16}$ \\
    \hline
    $ H^{a \alpha}, H_{a \alpha} $ & $ 2 \times 10^{16}$ \\
    \hline
    $ H_{ab(s)}, H^{ab}_{(s)} $ & $1 \times 10^{16}$ \\
    \hline
    $H^{a \alpha}_{\beta},H_{a \alpha}^{\beta}$ & $2 \times 10^{16}$ \\
    \hline
    $H^a,\overline{H}_a $ & $1 \times 10^{16}$ \\
    \hline
  \end{tabular}
  \caption{The Higgs spectrum for $aA=bA=cA=dA=a'A=b'A=c'A=d'A=2.4\times10^{16}$ GeV, $M'=M_1=M_1'=N=N'=2.4\times 10^{18}$ GeV, $M=2\times 10^{15}$ GeV, $M_V=M_{\Sigma}=2.4\times10^{16}$ GeV, $\Delta=10^6$ GeV and $P_1\equiv \frac{z_1 A}{M}=-3, P_2\equiv \frac{z_2 A}{M}=0$. For these parameters, $M_{c}=1.6 \times 10^{18}$ GeV, $\alpha_5(\Lambda)=\frac{1}{22}$, $  A=2.46\times 10^{16}$ GeV, $\lambda=0.195$, $\frac{\tilde{m}}{m}=0.4$, $\zeta=0.00098$ and $\eta=1$.}\label{5.9}
\end{table}

\section{Unification and Threshold Effects}\label{e}

At the GUT scale, the gauge coupling unification requires
\begin{eqnarray}
       &&(3\alpha_2^{-1}-2\alpha_3^{-1}-\alpha_1^{-1})(m_z)=\frac{1}{2 \pi}\Bigg\{-2\ln{\frac{m_{SUSY}}{m_z}}
       +\frac{6}{5}\ln{\frac {|\det{M^{eff}_T}|^2M_{H^{ab}_{(s)}}^9  M_{H^{a\alpha}_b}^4 M_{H^{a\alpha}}^4 M_{H^a}^7  }{m_z^2 M_+^2M_-^2 M_{H^{a\alpha}_{\beta}}^{24}} }\Bigg\},\label{5.1a} \\
       &&(5\alpha_1^{-1}-3\alpha_2^{-1}-2\alpha_3^{-1})(m_z)=\frac{1}{2 \pi}\Bigg\{8\ln{\frac{m_{SUSY}}{m_z}}+ 6\ln{\frac {M_V^4 M_{H^{ab}_{(s)}}M_{H^{a\alpha}_b}^4 M_{H^{a\alpha}_{\beta}}^6 M_{\Sigma}^2 }{m_z^6M_{H^{a\alpha}}^6   M_{H^a}^5}}\Bigg\}\label{5.1b}
\end{eqnarray}
at 1-loop level, where $M_V =\sqrt{\frac{5}{3}}~g_5 A $ is the mass of the $X,Y$ gauge superfields and $M_{\Sigma}=5 \lambda A$. The $\ln{\frac{m_{SUSY}}{m_z}}$-terms account for the  SUSY scale effects\cite{murayama1993}. At 2-loop level, the corrections
\begin{eqnarray}
       \delta_1^{(2)} & =& -\frac{1}{4\pi}\sum_{j=1}^3 \frac{1}{b_j} (3b_{2j}-2b_{3j}-b_{1j})\ln{\frac{\alpha_j(m_z)}{\alpha_5(\Lambda)}}, \label{5.2a}\\
       \delta_2^{(2)} & =&-\frac{1}{4\pi}\sum_{j=1}^3 \frac{1}{b_j} (5b_{1j}-3 b_{2j}-2 b_{3j})\ln{\frac{\alpha_j(m_z)}{\alpha_5(\Lambda)}}, \label{5.2b}
\end{eqnarray}
should be added on the r.h.s. in (\ref{5.1a}) and (\ref{5.1b}), respectively, where
\begin{equation}\label{5.3}
    b_i = \begin{pmatrix}
        \frac{33}{5} \\
        1\\
        -3
    \end{pmatrix},
    ~~b_{ij} = \begin{pmatrix}
        \frac{199}{25} & \frac{27}{5} & \frac{88}{5} \\
        \frac{9}{5} & 25 & 24\\
        \frac{11}{5} & 9 & 14
    \end{pmatrix}
\end{equation}
are the $\beta$-functions of the gauge couplings in the MSSM at 1- and 2-loop level, respectively. The number $\alpha_5(\Lambda)$ in (\ref{5.2b}) can take  its value at 1-loop level approximately.
% The threshold corrections at the two-loop level are expected to be small and are thus omitted.
Substituting the masses in Table \ref{3.11} into (\ref{5.1a}) and (\ref{5.1b}),
we have
\begin{eqnarray}
     &&(3\alpha_2^{-1}-2\alpha_3^{-1}-\alpha_1^{-1})(m_z)=\frac{1}{2 \pi}\Big\{-2\ln{\frac{m_{SUSY}}{m_z}}
     +\frac{12}{5}\ln{\frac {\zeta M_{c}}{m_z}}\Big\}+\delta_1^{(2)},\label{5.6a}\\
     &&(5\alpha_1^{-1}-3\alpha_2^{-1}-2\alpha_3^{-1})(m_z)=\frac{1}{2 \pi}\Big\{ 8\ln{\frac{m_{SUSY}}{m_z}}
     + 12\ln{\frac {\eta^3 \Lambda^3}{m_z^3 }}\Big\}+\delta_2^{(2)},\label{5.6b}
\end{eqnarray}
where
 \begin{eqnarray}
       && \zeta =\frac{|1+2P_1+2P_2|^{4.5}|1+2P_1-0.5P_2|^{2}|1-3P_1+2P_2|^{2}|1+2P_1-3P_2|^{3.5}} {|1-3P_1-0.5P_2|^{12}},\label{5.7a} \\
       && \eta^3 =\frac{|1+2P_1+2P_2|^{0.5}|1+2P_1-0.5P_2|^{2}|1-3P_1-0.5P_2|^{3}} {|1-3P_1+2P_2|^{3}|1+2P_1-3P_2|^{2.5}}\label{5.7b}
\end{eqnarray}
measure the mass splitting in $45+\overline{45}$, and $P_1=\frac{z_1 A}{M},P_2=\frac{z_2 A}{M}$.
The GUT scale is chosen as
\begin{equation}\label{5.5}
\Lambda=[M_V^2 M_{\Sigma}]^{\frac{1}{3}}=\big[\frac{25}{3}g_5^2 \lambda \big]^{\frac{1}{3}}A,
\end{equation}
and the ETM is
\begin{equation}\label{5.4}
M_{c}=\frac {|\det{M_T^{eff}}|}{M_+ M_-}.
\end{equation}

The SUSY scale effects are included by
taking into account the effects of top quark\cite{Yamada1993}, the mass splitting at SUSY scale\cite{murayama1993}, and the difference between the $\overline{MS}$-scheme and $\overline{DR}$-scheme\cite{Matin1993}. One can get the constraints\cite{murayama2002}
\begin{eqnarray}
     && 3.5\times 10^{14} GeV  \leq \zeta M_{c} \leq 3.6 \times 10^{15} GeV,\label{5.8a}\\
     && 1.7 \times 10^{16} GeV  \leq \eta \Lambda  \leq 2.0 \times 10^{16} GeV.\label{5.8b}
\end{eqnarray}
Note that a small $\zeta$ can enhance $M_{c}$ to suppress proton decay.

Using the  parameters in getting the typical spectrum  in Table \ref{5.9},  we plot the running  gauge couplings in Fig. \ref{5.10}. We can see that  perturbative region of the model is extended to $ \sim  10^{18}$ GeV, very close to the Planck scale, as less particles exist at the GUT scale  than those in the models with 75 breaking GUT symmetry.
\begin{figure}
\centering
       \includegraphics[scale= 0.95]{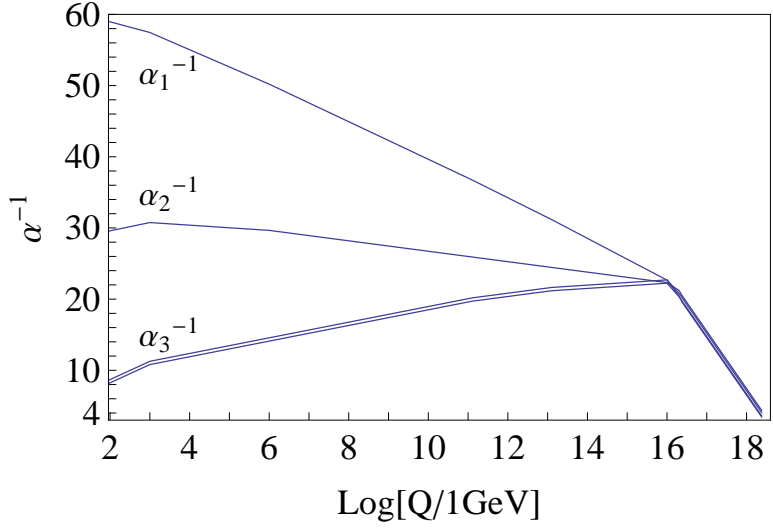}
       \caption{For those parameters used in Table \ref{5.9}, the running behaviors of the gauge  couplings.}\label{5.10}
\end{figure}

\section{Proton Decay and Constraints}\label{f}

The renormalizable couplings of the matter and Higgs  superfields are contained in the superpotential
\begin{equation}\label{6.1}
    W_F=\sqrt{2} f_1^{ij} \psi_i^{\alpha\beta} \phi_{j\alpha}\overline{5}_\beta+\sqrt{2} f_2^{ij} \psi_i^{\alpha\beta} \phi_{j\gamma}\overline{45}_{\alpha\beta}^\gamma+
    \frac{1}{4}h^{ij}\epsilon_{\alpha\beta\gamma\delta\epsilon}\psi_i^{\alpha\beta}
    \psi_j^{\gamma\delta}{5}^\epsilon,
\end{equation}
where the $U(1)_S$ and $U(1)_P$ quantum
numbers of $\psi_i$ and $\phi_i$  are chosen as $(0,0)$ and $(-1,-5)$, respectively.
Using the same parametrization as in \cite{zheng2012}, all the fermion masses and mixing can be generated correctly.
The matter fields are $\psi_i \ni (e^{-i\varphi_i} u_i^c, ~u_i, ~V_{ij}d_j, ~V_{ij}e_j^c )$ and $\phi_i \ni (d_i^c,~ \nu_i,~e_i)$. The corresponding coefficients are
\begin{equation}\label{}
    \begin{split}
      &h^{ij}= e^{i\varphi_i}\delta^{ij}\frac{m_{u_i}}{v_u},\\
      &f_1^{ij}=\frac{ V_{ij}^* M_+}{4 m}\big(\frac{3 m_{d_j}}{v_d}+\frac{m_{e_j}}{v_d}\big),\\
      &f_2^{ij}=\frac{ V_{ij}^* M_+}{2 a A}\big(\frac{m_{d_j}}{v_d}-\frac{m_{e_j}}{v_d}\big),
    \end{split}
\end{equation}
where two of the three phases $\varphi_i$'s are independent. Here $v_u$ and $v_d$ are the VEVs of the MSSM Higgs doublets.

Following \cite{Goto1999} the dominant mechanism of proton decay is through the wino dressed dimension-5 operators of the LLLL-type for $p \rightarrow K^++\overline{\nu}_{\mu(e)}$ and $p \rightarrow \pi^+ +\overline{\nu}_{\mu(e)}$, and through the higgsino dressed dimension-5 operators of the RRRR-type for $p \rightarrow K^++\overline{\nu}_{\tau}$ and $p \rightarrow \pi^+ +\overline{\nu}_{\tau}$.
The dimension-5 operators mediating proton decay are now
\begin{equation}\label{6.2}
W_5=C_{ijkl}(Q_iQ_j)(Q_kL_l)+ ~D_{ijkl}(u^c_i e^c_j)(u^c_k d^c_l),
\end{equation}
where by direct calculations
\begin{equation}\label{6.3}
    C_{ijkl}=\frac{1}{2}h^{ij}\big[f_1^{kl}({M_T^{eff}}^{-1})_{11}
    -f_2^{kl}({M_T^{eff}}^{-1})_{12}\big]
\end{equation}
for the LLLL  operators, and
\begin{equation}\label{6.4}
    D_{ijkl}=h^{im}V_{mj}e^{-i(\varphi_i+\varphi_k)}\big[f_1^{kl}({M_T^{eff}}^{-1})_{11}
    +f_2^{kl}({M_T^{eff}}^{-1})_{12}\big]
\end{equation}
for the RRRR operators.
From (\ref{4.9}),
\begin{equation}\label{6.5}
     ({M_T^{eff}}^{-1})_{11}=\frac{\tilde{m} \Delta}{\det{M^{eff}_T}},~~~~~~~~
      ({M_T^{eff}}^{-1})_{12}=\frac{a A \Delta}{\det{M^{eff}_T}}.
\end{equation}
then
\begin{eqnarray}
      && C_{ijkl} = \frac{ e^{i\varphi_i}\delta^{ij}}{ 2 M_{c}} V_{kl}^*\frac{m_{u_i}}{v_u} \Bigg[\Big(\frac{3}{4}\frac{\tilde{m}}{m}-\frac{1}{2}\Big)\frac{m_{d_l}}{v_d}
      +\Big(\frac{1}{4}\frac{\tilde{m}}{m}+\frac{1}{2}\Big)\frac{m_{e_l}}{v_d}\Bigg],\label{6.6a}\\
      && D_{ijkl} = \frac{e^{-\varphi_k}}{ 2 M_{c}}V_{ij} V_{kl}^* \frac{m_{u_i}}{v_u} \Bigg[\Big(\frac{3}{4}\frac{\tilde{m}}{m}+\frac{1}{2}\Big)\frac{m_{d_l}}{v_d}
      +\Big(\frac{1}{4}\frac{\tilde{m}}{m}-\frac{1}{2}\Big)\frac{m_{e_l}}{v_d}\Bigg],\label{6.6b}
\end{eqnarray}
which show that  $M_{c}$, the ETM,  determines these coefficients and  thus proton lifetime. Note that $\frac{\tilde{m}}{m}= \frac{1-0.5P_1+0.75P_2}{1-1.75P_1-2.375P_2}$  depends only on $P_1$ and $P_2$.
The main partial lifetimes  are given in Table \ref{6.7} which give constraints on $P_1$, $P_2$ and $M_c$ by comparing with the data\cite{pdg}. Combining with the bound (\ref{5.8a}), it gives constraints on parameters $P_1$, $P_2$.

In the present model $\alpha_5 (\Lambda) \sim \frac{1}{22}$,
 so $ \Lambda \sim 1.68 \lambda^\frac{1}{3} A$ following (\ref{5.5}). Varying $\lambda$ in $0.1\sim 1$ and  A in $1 \times 10^{16} \sim 4 \times 10^{16}$ GeV, then from (\ref{5.8b}), we have
\begin{equation}\label{6.9}
    0.25<{\eta}<2.56,
\end{equation}
which gives another constraint on $P_1$ and $P_2$. Combining with these constraints, we can get the excluded region for the parameters $P_1$ and $P_2$,  which is plotted in Fig. \ref{6.10}. We can see that the parameter space is largely constrained.
\begin{table}
 \centering
  \begin{tabular}{|c|c|c|c|}
    \hline
    \multirow{2}{*} {Decay mode} & Partial lifetime  & Typical number& Data\cite{pdg}\\
     & (years) & (years) & (years)\\
    \hline
    $ \tau(p \rightarrow \overline{\nu}_\mu ~\pi^+)$  & $\frac{1.38}{(0.25 + 10.125 \tilde{m}/{m})^2} (\frac{M_c}{GeV})^2$ &$ 1.9\times10^{35} $ & \multirow{3}{*} {$ >2.5 \times 10^{31} $} \\
     \cline{1-3}
    $ \tau(p \rightarrow \overline{\nu}_e ~\pi^+)$  & $\frac{7.366}{(2.25 - 3.875 \tilde{m}/{m})^2} (\frac{M_c}{GeV})^2$ &$ 3.8\times 10^{37}  $  & \\
    \cline{1-3}
    $ \tau(p \rightarrow \overline{\nu}_\tau ~\pi^+)$  & $\frac{1.04}{(1.761- 5.984 \tilde{m}/{m})^2} (\frac{M_c}{GeV})^2$ &$ 6.6\times10^{36}  $  & \\
     \hline
    $\tau(p \rightarrow \overline{\nu}_\mu ~K^+)$ & $\frac{0.664}{(0.25 + 10.125 \tilde{m}/{m})^2}  (\frac{M_c}{GeV})^2$ &$ 9.2 \times 10^{34}  $  & \multirow{3}{*} {$>6.7 \times 10^{32}$}  \\
      \cline{1-3}
    $\tau(p \rightarrow \overline{\nu}_e ~K^+)$ & $\frac{5.712}{(2.858 - 4.921 \tilde{m}/{m})^2} (\frac{M_c}{GeV})^2$ &$ 1.9 \times 10^{37}  $  &  \\
     \cline{1-3}
    $\tau(p \rightarrow \overline{\nu}_\tau ~K^+)$ & $\frac{1.904}{(3.152 - 12.555 \tilde{m}/{m})^2} (\frac{M_c}{GeV})^2$ &$ 1.4 \times 10^{36} $ & \\
    \hline
  \end{tabular}
  \caption{The calculated proton partial lifetimes vs data. The typical numbers are got by taking the representative parameters in Table \ref{5.9}.}\label{6.7}
\end{table}

\begin{figure}
\centering
       \includegraphics[scale= 0.85]{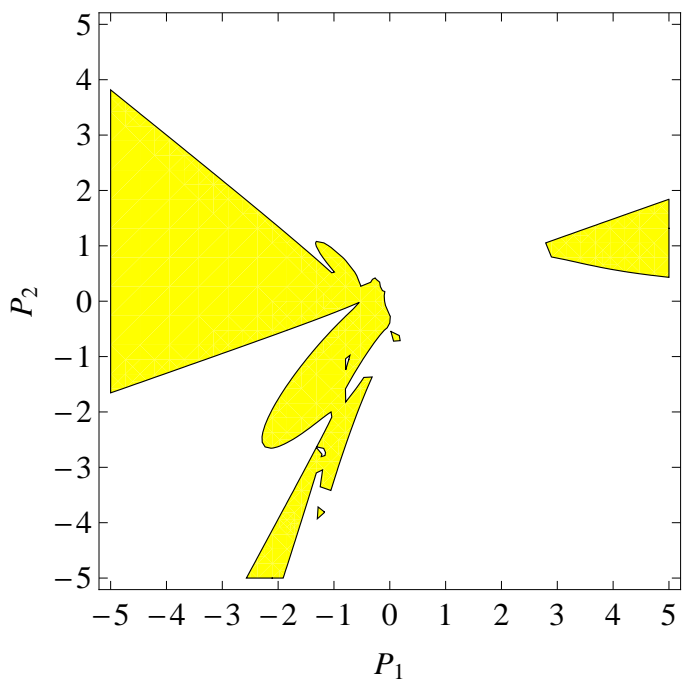}
       \caption{Constraints on the parameters $P_1$ and $P_2$.
The blank region is excluded by gauge coupling unification and proton partial lifetimes.}\label{6.10}
\end{figure}

\section{Summary}\label{g}

We have presented a realistic MPM using 24-plet instead of 75-plet to break the GUT symmetry.
This model is renormalizable at a high scale close to the Planck scale.
At the GUT scale the model contains one $24$-plet, one singlet, two pair of $5+\overline{5}$s, one pair of $45+\overline{45}$ in the Higgs sector. Below the GUT scale with the $U(1)_S$ symmetry breaking effects,
the Higgs spectrum  contains a pair of weak doublets at $10^6$GeV in addition to the MSSM doublets, and two
pairs of color triplets at $10^{11-13}$GeV.

This model uses the GJM to account for the fermion masses and mixing. The GUT gauge coupling
remains to be perturbative even close to the Planck scale. The ETM is enhanced and proton decay is suppressed.

This work was supported in part by the National Natural Science Foundation of China (NSFC) under Grant No. 10435040.

\end{document}